\pdfoutput=1
\documentclass[11pt]{article}
\usepackage[]{acl}
\usepackage{times}
\usepackage{url}
\usepackage{color}
\usepackage{latexsym}
\usepackage{graphicx}
\usepackage{subfigure}
\usepackage{xspace}
\usepackage{arydshln}
\usepackage{booktabs}
\usepackage{multirow}
\usepackage{stfloats}
\usepackage{hyperref}
\usepackage{amsmath,amssymb,mathrsfs,utfsym,bm}

\usepackage[T1]{fontenc}
\usepackage[utf8]{inputenc}

\usepackage{microtype}
\newcommand{\adcap}{\textsc{AdPool}\xspace}
\newcommand{\adcto}{\textsc{AdOpt}\xspace}

\title{Improving Visual-Semantic Embedding with Adaptive Pooling and Optimization Objective}
\author{
    \\ Zijian Zhang$^{*, 1}$, Chang Shu\thanks{$\quad$  These authors contributed equally to this work.} $^{ ,2,3}$, Ya Xiao$^{1}$, Yuan Shen$^{1}$, Di Zhu$^{1}$, Jing Xiao$^{2}$, \\
    Youxin Chen$^{2}$, Jey Han Lau$^{4}$, Qian Zhang$^{3}$ and Zheng Lu$^{3}$ \\
    $^{1}$ Meituan, China \\ 
    $^{2}$ Ping An Technology (Shenzhen) Co., Ltd, China \\
    $^{3}$ University of Nottingham Ningbo, China \\
    $^{4}$ The University of Melbourne, Australia}

\begin{document}
\maketitle
\begin{abstract}
Visual-Semantic Embedding (VSE) aims to learn an embedding space where related visual and semantic instances are close to each other. Recent VSE models tend to design complex structures to pool visual and semantic features into fixed-length vectors and use hard triplet loss for optimization. However, we find that: (1) combining simple pooling methods is no worse than these sophisticated methods; and (2) only considering the most difficult-to-distinguish negative sample leads to slow convergence and poor Recall@K improvement. To this end, we propose an adaptive pooling strategy that allows the model to learn how to aggregate features through a combination of simple pooling methods. We also introduce a strategy to dynamically select a group of negative samples to make the optimization converge faster and perform better. Experimental results on Flickr30K and MS-COCO demonstrate that a standard VSE using our pooling and optimization strategies outperforms current state-of-the-art systems (at least 1.0\% on the metrics of recall) in image-to-text and text-to-image retrieval. Source code of our experiments is available at \texttt{\url{https://github.com/96-Zachary/vse_2ad}}.
\end{abstract}

\section{Introduction}
\label{intro}
\begin{figure}[t]
\centering
\includegraphics[width=0.9\linewidth]{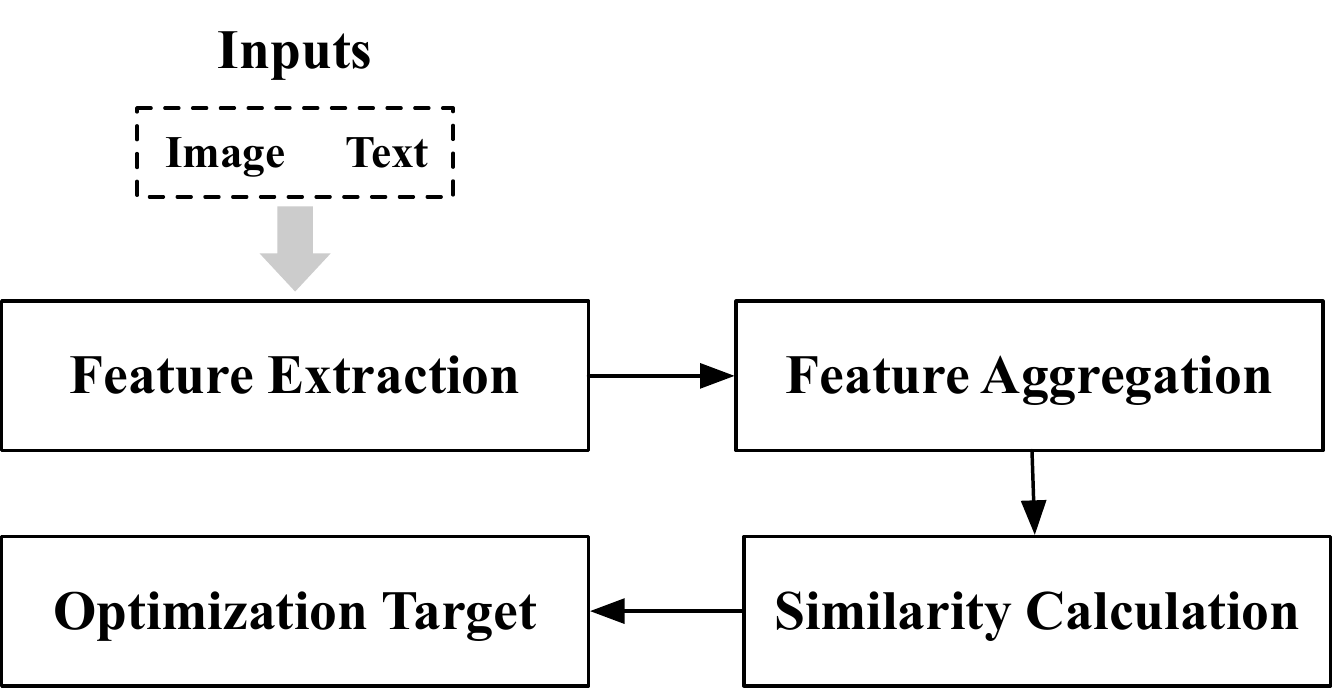}
\caption{\textbf{Illustration of VSE}. \label{fig:vse_steps}}
\end{figure}

Visual Semantic Embedding (VSE) \cite{frome2013devise,faghri2018vse++} is representation learning method that embeds images and texts for efficient cross-modal retrieval, and typically has the following steps (see Figure~\ref{fig:vse_steps} for an illustration). The image and text are extracted as features by separate visual and text encoders. These features are then projected into a joint embedding space and pooled to form fixed-length vectors. Then, the similarity calculation is utilized to measure the distance between instances and a suitable target is chosen for optimization. Our paper focuses on improving the steps of feature aggregation and optimization.

For feature aggregation, the most commonly used methods are simple pooling aggregators. MaxPool \cite{wang2018learning} and MeanPool \cite{reimers-gurevych-2019-sentence} are designed to detect the salient and mean points of features, and K-MaxPool \cite{kalchbrenner-etal-2014-convolutional} extracts the mean of top-${\rm K}$ features. Some complex aggregation techniques have been proposed, e.g.\ local-importance projection \cite{9010729}, sequence-to-sequence encoder \cite{hu2019evaluating}, graph convolution network \cite{li2019visual}, exponential adaptive pooling \cite{stergiou2021adapool} and self-attention encoder \cite{wang2020consensus}. However, we found that carefully selected pooling functions can surpass complex methods (see Appendix \ref{verify_a1}). Motivated by this, our paper proposes an approach that can \emph{automatically} discover the best pooling functions. Specifically, we seek to improve the feature aggregation step by proposing a formulation that parameterizes the different pooling strategies and allows the model to learn the best configuration automatically via its objective, alleviating the need to do manual tuning. In other words, we've turned these hyper-parameters (i.e.\ choices of pooling functions) into parameters in the model.

For optimization, most VSE models optimize using the hinge triplet ranking loss with in-batch negative example \cite{faghri2018vse++}. The intuition of the objective is to encourage positive pairs to be embedded in a similar space while widening the distance between a target with the hardest in-batch negative sample. In practice, however, it is often difficult for the model to find a good negative sample in the early stages of training (as instances are randomly distributed in space), resulting in slow convergence (see Appendix \ref{verify_a2}). To improve optimization, we propose an adaptive optimization objective that selects multiple in-batch negative samples based on model quality during training. The intuition is that in the early stages of training we want to sample more negative samples, and in the later stages fewer negative samples. 

Over two public datasets, MS-COCO \cite{lin2014microsoft} and Flickr30K \cite{young2014image}, we show that a standard VSE model using our proposed feature aggregation and optimization strategies outperforms benchmark models substantially.  In particular, our method obtains 1.4\% relative gains on RSUM for MS-COCO and 1.0\% for Flickr30K.  Compared with the pre-trained vision-language model with similar performance \citet{tacl_a_00473}, our method is ${\rm 4.3\times}$ faster.


\section{Related Work}
\label{related-work}
Depending on whether the image and text features have any form of \emph{cross-modal interaction} before similarity calculation, existing image-text retrieval can be broadly categorized into two types. 

The visual semantic embedding (VSE) \cite{faghri2018vse++, wang2020consensus, chun2021probabilistic} methods process the multimodal instances \emph{independently} before projecting them into a joint embedding space for similarity matching. \citet{wang2018learning} design a two-branch neural networks, LIWE \cite{wehrmann2019language} considers character-based alignment and embedding methods for language encoder and \citet{faghri2018vse++} extend it by using hard triplet loss for optimization. Following these ideas, PVSE \cite{song2019polysemous} and CVSE \cite{wang2020consensus} are proposed to consider intra-modal polysemous and consensus information. Recently, \citet{chun2021probabilistic} samples instances as probabilistic distributions and achieves further improvement. These VSE-based methods are fast as they do not consider cross-modal interaction and as such the visual and text features can be pre-computed. The non-VSE methods concentrate on the interaction of modalities. Specially, late-interaction methods explore to fusion multi-modal information by attention \cite{lee2018stacked, chen2020imram}, alignment \cite{zhang2020context}, multi-view representation \cite{qu2020context} and fine-grained reasoning \cite{qu2021dynamic}. The early-interaction methods \cite{tacl_a_00473}, like pre-trained vision-language models \cite{lu2019vilbert, chen2019uniter, li2020oscar, Jia2021ScalingUV, li2022mvp}, focuses on the maximum of performance while sacrifices efficiency.


Our paper focuses on the improvement of feature aggregation and optimization for VSE. The existing explorations of those two steps are as follows.

The performance of VSE ultimately depends on the quality of the joint embedding space, which is usually learned with simple transformations (e.g.\ linear projection or multi-layer perceptron) and pooling aggregators (e.g.\ mean pooling \cite{faghri2018vse++, qu2020context}, max pooling \cite{zhang2018deep, li2021memorize}, or a combination of them \cite{lee2018stacked}). Compared to these simple aggregation methods, more complex aggregators that introduce a large number of trainable parameters have also been explored, e.g.\ inter-modal attention \cite{wehrmann2020adaptive} and self-attention mechanisms \cite{han2021fine}. \citet{zhang2021heterogeneous} design a cross-modal guided pooling module that attends to local information dynamically. These sophisticated aggregators typically require more time, and don't always outperform simple pooling strategies. Perhaps the closest study to our work is GPO (VSE$\infty$) \cite{chen2021learning}, which builds a generalized operator to learn the best pooling strategy that only considers the position information of the extracted features.

Some studies focus on improving the optimization objective, and the most widely adopted objective is the hinge-based hard triplet ranking loss \cite{faghri2018vse++, wei2020multi, messina2021transformer}, which dynamically selects the ``hardest'' negative sample within a mini-batch. Other studies explore solutions that choose multiple negative samples. \citet{zhou2020ladder} introduce a coherence metric to rank the ``irrelevant'' candidates. Extending the idea, \citet{wei2020universal} assign different weights for positive and negative pairs. To tackle the issue of noisy labels which impacts multimodal representation,  \citet{hu2021learning} propose maximizing the mutual information between different modalities. \citet{huang2021learning} separate data into ``clean'' and ``noisy'' partitions by co-teaching. However, the above methods do not change adaptively according to the model performance when selecting negative samples.

\section{Methodology}
\label{optim}
\begin{figure*}[t]
\centering
\includegraphics[width=\linewidth]{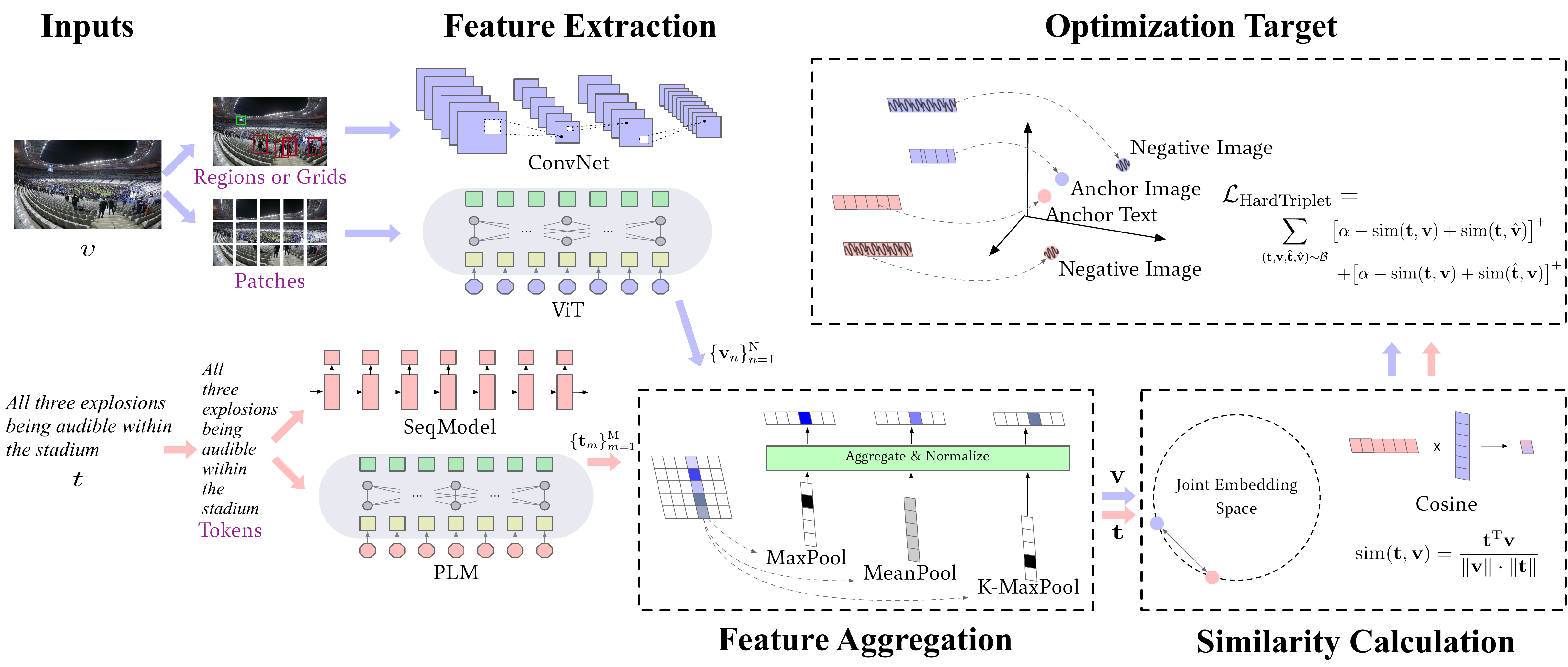}
\caption{\textbf{The framework of VSE.} The visual and text encoders process the image and text separately at first. The related images and sentences are then directed to a similar space using an appropriate optimization target. \label{fig:standard}}
\end{figure*}

\subsection{Background of VSE}
\label{standvse}
We first discuss the standard formulation of VSE, before introducing our innovation on improving feature aggregation (Section \ref{sec:pooling}) and optimization (Section \ref{sec:optimization}).

To compute the similarity of given multimodal instance (image \& text), a VSE model (Figure~\ref{fig:standard}) separately encodes them via a visual encoder ($\mathtt{VisEnc}(\cdot)$) and a text encoder ($\mathtt{TextEnc}(\cdot)$). There are three widely used visual features produced by different visual encoders --- \emph{grid} is the feature maps from convolutional networks (CNNs; \citet{he2016deep}), \emph{region} is the region of interest features from object detectors \cite{anderson2018bottom} and \emph{patch} is the partition from vision transformer \cite{dosovitskiy2020vit}. The text encoders are usually RNNs \cite{sutskever2014sequence}) and BERT \cite{devlin-etal-2019-bert}. Formally:
\begin{equation*}
    \begin{aligned}
        \textbf{F}_v &= \mathtt{VisEnc}(v)\\
        \textbf{F}_t &= \mathtt{TextEnc}(t) \\
    \end{aligned}
\end{equation*}
where $v$ and $t$ are the input image and text.

Assuming the visual feature $\textbf{F}_v$ has ${\rm N}$ object vectors (represented either as grids, regions or patches) in $d_1$ dimension, and the text feature $\textbf{F}_t$ has ${\rm M}$ token vectors in ${d_2}$ dimension, we next project them to the same $d$-dimension:

\begin{equation}
    \begin{aligned}
        \{\textbf{v}_n\}^{\rm N}_{n=1} = \textbf{F}_v\textbf{W}_v + \textbf{b}_v \\ \{\textbf{t}_m\}^{\rm M}_{m=1} = \textbf{F}_t\textbf{W}_t + \textbf{b}_t
    \end{aligned}
\label{eqn:features}
\end{equation} 
where $\mathbf{v}_n$ and $\mathbf{t}_m$ now have the same dimension $d$.

To aggregate the extracted features into fixed-length vectors, domain aggregators, $f_{\rm vision}(\cdot)$ and $f_{\rm text}(\cdot)$ are used to transform $\{\textbf{v}_n\}^{\rm N}_{n=1} \in \mathbb{R}^{{\rm N} \times d}$ and $\{\textbf{t}_n\}^{\rm M}_{m=1} \in \mathbb{R}^{{\rm M} \times d}$ into $\textbf{v} \in \mathbb{R}^d$ and $\textbf{t} \in \mathbb{R}^d$, respectively: 
\begin{equation*}
    \begin{aligned}
        \textbf{v} &=f_{\rm vision}\Big(\{\textbf{v}_n \}^{\rm N}_{n=1}\Big),\\
        \textbf{t} &=f_{\rm text}\Big(\{\textbf{t}_m \}^{\rm M}_{m=1}\Big)
    \end{aligned}
\end{equation*} 

And lastly, to measure how related the inputs we use cosine similarity:
\begin{equation*}
        {\rm sim}(\textbf{t}, \textbf{v}) = \frac{\textbf{t}^{\mathrm{T}} \textbf{v}}{\Vert \textbf{t} \Vert \cdot \Vert \textbf{v} \Vert}
\end{equation*}


Existing optimization strategies generally use the hinge-based triplet ranking loss to optimize the VSE model. Given an anchor, it aims to maximize its similarity with positive samples while minimizing its similarity with the most ``difficult'' negative sample in the mini-batch (i.e.\ the example that has the highest similarity with the anchor that is not a positive example), and includes both text-to-image and image-to-text retrieval objectives:
\begin{equation}
    \begin{aligned}
        \mathcal{L}_{\rm HardTriplet} &= \\ 
        \sum_{(\textbf{t}, \textbf{v}, \hat{\textbf{t}}, \hat{\textbf{v}}) \sim \mathcal{B}} &\big[\alpha - {\rm sim}(\textbf{t}, \textbf{v}) + {\rm sim}(\textbf{t}, \hat{\textbf{v}})\big]^{+} \\
        &+\big[\alpha - {\rm sim}(\textbf{t}, \textbf{v}) + {\rm sim}(\hat{\textbf{t}}, \textbf{v}) \big]^{+}
    \end{aligned}
\label{hard-triplet}
\end{equation} 
where $\alpha$ is the margin hyper-parameter, and $[x]^+=\mathrm{max}(0,x)$. $(\textbf{t}, \textbf{v})$ is a positive text-image pair in mini-batch $\mathcal{B}$ and $(\hat{\textbf{t}}, \textbf{v})$ and $(\textbf{t}, \hat{\textbf{v}})$ are negative pairs, where $\hat{\textbf{t}}=\mathrm{argmax}_{\textbf{t}'\neq\textbf{t}} {\rm sim}(\textbf{t}',\textbf{v})$ and $\hat{\textbf{v}}=\mathrm{argmax}_{\textbf{v}'\neq\textbf{v}} {\rm sim}(\textbf{t},\textbf{v}')$ are the hardest negative sentence and image respectively in $\mathcal{B}$.

\subsection{Enhancing VSE by Adaptive Pooling}
\label{sec:pooling}
We now describe our approach (named adaptive pooling, \adcap) to improving feature aggregation, where we parameterize the pooling operations ($f_{\rm vision}$ and $f_{\rm text}$) to allow VSE to learn the best way to aggregate the features via its objective adaptively (in other words, we have effectively turned the pooling methods --- which are hyper-parameters --- into model parameters). Our approach has the advantage of being much faster and simpler than complex aggregators, as it introduces only a small number of parameters. Note that we will describe our method only for the text feature ($\{\textbf{t}_m \}^{\rm M}_{m=1}$), as the same process can be applied for the visual feature ($\{\textbf{v}_n\}^{\rm N}_{n=1}$).

\subsubsection{Token-level Pooling}
\label{sec:object-pooling}
\begin{figure*}[t]
\centering
\includegraphics[width=\linewidth]{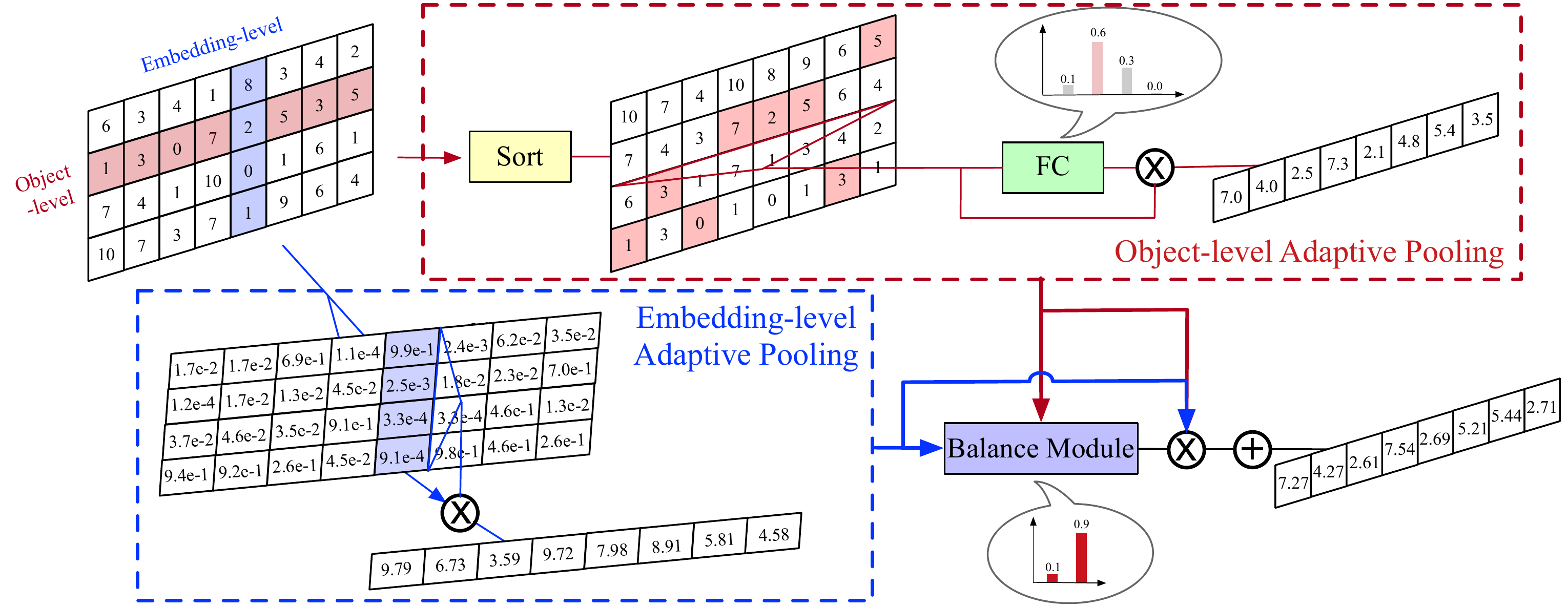}
\caption{\textbf{Illustration of our proposed adaptive pooling module (\adcap).} It produces vectors pooled at the Token-level and embedding-level and combined via a balance module. Each red row denotes a token in a sentence.
\label{fig:adcap}}
\end{figure*}

Recall that the text feature ($\{\textbf{t}_m \}^{\rm M}_{m=1}$) has ${\rm M}$ tokens each of $d$-dimension. The exact form of $\textbf{t}_m$ can be token vectors. As for vision features, they can be grids, regions or image patches, depending on the visual encoder. Let $\mathrm{max}_k(\cdot)$ be a function that extracts the top $k$-th value in a list, the common pooling mechanisms can be formulated as a \emph{"sort $\rightarrow$ weight-sum"} paradigm \cite{grefenstette2014convolutional} as following:
\begin{itemize}
\setlength{\itemsep}{0pt}
\setlength{\parsep}{0pt}
\setlength{\parskip}{0pt}
    \item \textbf{MeanPool} = $\frac{1}{\rm M}\sum^{\rm M}_{i=1}\{{t}_{ij}\}^{\rm M}_{i=1}$
    ouputs the mean value among ${\rm N}$ objects at position $j$; 
    
    \item \textbf{MaxPool} = $\mathrm{max}_1\big(\{{t}_{ij}\}^{\rm M}_{i=1} \big)$ 
     returns the maximum values of $\{{t}_{ij}\}^{\rm M}_{i=1}, \forall{j}$; 
     
    \item \textbf{K-MaxPool} = $\frac{1}{\rm K}\sum^{\rm K}_{k=1}\mathrm{max}_k\big(\{{t}_{ij}\}^{\rm M}_{i=1} \big)$ computes the mean of the top-${\rm K}$ values.
\end{itemize}
where $t_{ij}$ denotes the $j$-th element in the $i$-th token vector.

Using the token vectors $\{\textbf{t}_m\}^{\rm M}_{m=1}$ in Figure~\ref{fig:adcap} (red rows in the input matrix) as an example, we follow the \emph{"sort $\rightarrow$ weight-sum"} paradigm of simple pooling methods to first sort the feature matrix along the embedding axis. Next we let the model learn (via a fully-connected layer) how to weight each vector (red rows in the matrix) automatically  (e.g.\ $[0.1,0.6,0.3,0.0]$ in Figure~\ref{fig:adcap}). In this way, we allow the model to find the best combination of MeanPool and K-MaxPool in an adaptive manner. Formally as:
\begin{equation}
    \begin{aligned}
        \textbf{t}^{\rm tok} &= \sum^{\rm M}_{m=1} \theta_m \cdot  \textbf{u}_m \\
        \{\textbf{u}_m \}^{\rm M}_{m=1} &= \mathrm{sort}\big(\{\textbf{t}_m \}^{\rm M}_{m=1}\big) \\
         \\
        \boldsymbol{\theta} &= \mathrm{softmax}\left(\{\textbf{u}_m \}^{\rm M}_{m=1} \textbf{W}_{\rm tok}\right)
    \end{aligned}
\end{equation}
where $\mathrm{sort(\cdot)}$ is a function that sorts the token vectors along the embedding axis.

\subsubsection{Embedding-level Pooling}
\label{sec:embedding-pooling}
With token-level pooling, we learn to how to weight each (sorted) token vector and aggregate them. With embedding-pooling, we learn how to weight each (original unsorted) object vector \textit{in each embedding dimension} to extract the most salient element within that dimension, and can be interpreted as a ``soft'' version of MaxPool:
\begin{equation}
    \begin{aligned}
        \textbf{t}^{\rm emb} &= \sum^{\rm M}_{i=1} \delta_{ij} \cdot {t}_{ij}, \forall{j}\\
        \delta_{ij} &= \frac{e^{{t}_{ij}}}{\sum^{\rm M}_{i=1}e^{{t}_{ij}}}, \forall{i}
    \end{aligned}
\end{equation}

\subsubsection{Combining Token-level and Embedding-level Pooled Vectors}
\label{sec:combine-module}
Given the two pooled vectors $\textbf{t}^{\rm tok}$ and $\textbf{t}^{\rm emb}$ (which have the same dimension $d$), one straightforward way to combine them is to weight them with scaling hyper-parameters. To avoid these hyper-parameters (which requires manual tuning), we let the model learn these weights automatically with a trainable $\textbf{W}_{bal} \in \mathbb{R}^{d \times 1}$:
\begin{equation}
    \begin{aligned}
        \textbf{t} &= \omega_1\textbf{t}^{\rm tok} + \omega_2\textbf{t}^{\rm emb}\\
        \mathbf{\omega_{\{1,2\}}} &= \text{softmax}\Big( [\textbf{t}^{\rm tok}\textbf{W}_{bal}, \textbf{t}^{\rm emb}\textbf{W}_{bal}]\Big)
    \end{aligned}
    \label{eqn:balance}
\end{equation}

For the visual feature $\{\textbf{v}_n \}^{\rm N}_{n=1}$ (Equation \ref{eqn:features}), we follow the same process to compute the token-level (where here an ``token'' is a unit of image) and embedding-level pooled vectors ($\textbf{v}^{\rm tok}$ and $\textbf{v}^{\rm emb}$) and combine them to produce $\textbf{v} \in \mathbb{R}^d$.

\subsection{Enhancing VSE by Adaptive Objective}
\label{sec:optimization}
Our next contribution is in improving the optimization step.
The hard triplet loss (Equation \ref{hard-triplet}) is the most commonly utilized training objective for optimizing VSE. However, we find that locating a ``difficult'' negative sample is challenging in the early stages of training (Appendix \ref{verify_a2}), resulting in delayed convergence. We introduce a novel adaptive optimization objective, \adcto, that automatically (or adaptively) selects ${\rm K}$ ${\rm (K\geq1)}$ negative samples in each mini-batch $\mathcal{B}$ during training.

\citet{wang2020understanding} introduce two key properties, \emph{Alignment} and \emph{Uniformity}, that correlate with a retrieval performance:
\begin{itemize}
\setlength{\itemsep}{1pt}
\setlength{\parsep}{0pt}
\setlength{\parskip}{0pt}
    \item \emph{Alignment}: the positive $(\textbf{t}$ and $\textbf{v})$ should be mapped into a similar embedding space. Defining $\gamma_{\rm align} \in [0,1]$ as the average similarity values for all positive pairs, the larger $\gamma_{\rm align}$ is, the better the VSE model.
    \begin{equation*}
        \gamma_{\rm align} = \mathop{\mathbb{E}}\limits_{(\textbf{t}, \textbf{v}) \sim \mathcal{B}}[{\rm sim}(\textbf{t}, \textbf{v})]
    \end{equation*}
    
    \item \emph{Uniformity}: all vectors should be roughly uniformly distributed on the unit hypersphere, and $\gamma_{\rm uniform} \in [0,1]$ measures this quality.
    \begin{equation*}
        \gamma_{\rm uniform} = \mathrm{log} \mathop{\mathbb{E}}\limits_{\textbf{t} \sim \mathcal{B}, \textbf{v} \sim \mathcal{B}}[ e^{{\rm sim}(\textbf{t}, \textbf{v})}]
    \end{equation*}
\end{itemize}

By combining $\frac{(\gamma_{\rm align}+\gamma_{\rm uniform}) \times \pi}{4} \in [0,\frac{\pi}{2}]$, we can use it to assess the maturity of a VSE model during training and dynamically adjust the number of negative samples. That is, in the early stages of training, we expect the value to be close to 0 (as the model is unable to differentiate between positive and negative samples), and we would want to use more negative samples for optimization. Conversely in the later stages of training, the value should be close to $\frac{\pi}{2}$ and we want less negative samples. Formally:
\begin{equation}
\begin{aligned}
    {\rm K}' &= \rm{round}\Big(|\mathcal{B}| \times \mathrm{cos}(\frac{(\gamma_{\rm align}+\gamma_{\rm uniform}) \times \pi}{4} )\Big) \\
    \rm{K} & = \rm{max}\Big(1, \rm{min}(\rm{K}', |\mathcal{B}| -1) \Big) \label{eqn:k}
\end{aligned}
\end{equation}
where $\rm{cos}$ is the cosine function to invert the sum\footnote{The purpose of using $\mathrm{cos}$ is to map $\frac{(\gamma_{\rm align}+\gamma_{\rm uniform}) \times \pi}{4} \in [0,\frac{\pi}{2}]$ to the range of $[1,0]$.}, $\rm{K}$ is the number of negative samples to be used, and ${\rm round}(x)$ is a function that rounds down the value of $x$.
As the hard triplet loss can only work with ${\rm K}=1$ (Equation \ref{hard-triplet}), we therefore utilize the InfoNCE objective \cite{van2018representation}, which is a commonly used contrastive objective \cite{radford2021learning}:

\begin{equation*}
    \begin{aligned}
        &\mathcal{L} = \mathcal{L}_{\rm v2t} + \mathcal{L}_{\rm t2v} \\
        &\mathcal{L}_{\rm v2t} = -\frac{1}{|\mathcal{B}|}\sum_{(\textbf{t}, \textbf{v})\in \mathcal{B}}\mathrm{log}\frac{\mathrm{exp\big(\mathrm{sim}(\textbf{t}, \textbf{v})/\tau \big)}}{\sum^{\rm K}_k \mathrm{exp}\big(\mathrm{sim}(\hat{\textbf{t}}_k, \textbf{v})/\tau \big)}\\
        &\mathcal{L}_{\rm t2v} = -\frac{1}{|\mathcal{B}|}\sum_{(\textbf{t}, \textbf{v})\in \mathcal{B}}\mathrm{log}\frac{\mathrm{exp\big(\mathrm{sim}(\textbf{t}, \textbf{v})/\tau \big)}}{\sum^{\rm K}_k \mathrm{exp}\big(\mathrm{sim}(\textbf{t}, \hat{\textbf{v}}_k)/\tau \big)}
    \end{aligned}
\end{equation*} 
where $\tau$ is the temperature hyper-parameter.

\section{Experiments}
\label{experiments}
\subsection{Datasets and Metrics}
\textbf{Datasets}. We conduct our experiments on MS-COCO and Flickr30K using various visual and text encoders for cross-modal retrieval. The MS-COCO dataset contains 123,287 images, each with 5 manually annotated sentences. Following the split method of \citet{faghri2018vse++}, 113,287 images are used for training, 5,000 for validation, and 5,000 for testing. 

Following prior studies \cite{faghri2018vse++}, we experiment with two evaluation settings: (1) MS-COCO 1K averages the results over 5-folds of 1K test images; and (2) MS-COCO 5K directly results on the whole 5K test images. Following \citet{chen2021learning}, we use the former to assess overall performance with state-of-the-art VSE models and the latter for further analyses such as ablation results. Flickr30K consists of 31,783 images with the same corresponding 5 captions, and 1,000 images are reserved for validation and testing.

\textbf{Metrics}. We evaluate cross-modal retrieval performance using recall@K (R@K), where ${\rm K}=\{1,5,10\}$. The evaluation tasks include both caption retrieval (given caption, find images) and image retrieval (given image, find captions). We also compute RSUM which is a sum of all R@K metrics across both tasks to assess the overall performance.

\begin{table*}[t]
\centering
\small{
\renewcommand\arraystretch{1.1}
\setlength\tabcolsep{1.5pt}
\begin{tabular}{clccc:ccc:c|}
\toprule
   &  & \multicolumn{6}{c}{MS-COCO 5-fold 1K Test} & \\ \cline{3-8} 
   &  & \multicolumn{3}{c}{Image Retrieval} & \multicolumn{3}{c}{Caption Retrieval} &  \\
{VF} & Method    & R@1  & R@5  & R@10  & R@1  & R@5  & R@10  & RSUM  \\ \hline
& \multicolumn{8}{l}{\textbf{Text Encoder}: BiGRU}\\ 
R &VSE++  & 54.0  & 85.6  & 92.7  & 68.5  & 92.6  & 97.1  & 490.5  \\
R &LIWE        & 57.9  & 88.3  & 94.5  & 73.2  & 95.5  & 98.2  & 507.6 \\
R &PVSE          & 55.2  & 86.5  & 93.7  & 69.2  & 91.6  & 96.6  & 492.8 \\
R &CVSE        & 55.7  & 86.9  & 93.8  & 69.2  & 93.3  & 97.5  & 496.4 \\
R &VSE$\infty$ & 61.7  & 90.3  & 95.6  & 78.5  & 96.0  & 98.7  & 520.5 \\
R &SCAN(i2t)   & 54.4  & 86.0  & 93.6  & 69.2  & 93.2  & 97.5  & 493.9 \\ 
R &SCAN(t2i)   & 56.4  & 87.0  & 93.9  & 70.9  & 94.5  & 97.8  & 500.5 \\
R &CAAN        & 61.3  & 89.7  & 95.2  & 75.5  & 95.4  & 98.5  & 515.6 \\
R &IMRAM       & 61.7  & 89.1  & 95.0  & 76.7  & 95.6  & 98.5  & 516.6 \\
R &VSE+2AD  & \textbf{63.5}  & \textbf{91.8}  & \textbf{96.3}  & \textbf{79.7}  & \textbf{97.3}  & \textbf{99.2}  & \textbf{527.8} \\ \cdashline{1-9}[1pt/1pt]
RG &VSE$\infty$  & 64.8  & 91.6  & 96.5  & 80.0  & 97.0  & 99.0  & 528.8  \\
RG &VSE+2AD       & \textcolor{blue}{\textbf{65.7}}  & \textcolor{blue}{\textbf{92.3}}  & \textcolor{blue}{\textbf{97.0}}  & \textcolor{blue}{\textbf{82.1}}  & \textcolor{blue}{\textbf{97.9}}  & \textcolor{blue}{\textbf{99.4}}  & \textcolor{blue}{\textbf{534.4}} \\ \cdashline{1-9}[1pt/1pt] 
RGP &VSE+2AD & \textcolor{red}{\textbf{67.1}}  & \textcolor{red}{\textbf{93.0}}  & \textcolor{red}{\textbf{97.7}}  & \textcolor{red}{\textbf{83.8}}  & \textcolor{red}{\textbf{98.1}}  & \textcolor{red}{\textbf{99.4}}  & \textcolor{red}{\textbf{539.1}} \\ \hline \hline
& \multicolumn{8}{l}{\textbf{Text Encoder}: BERT}   \\ 
R &VSE++  & 54.0  & 85.6  & 92.5  & 67.9  & 91.9  & 97.0  & 488.9 \\
R &VSE$\infty$  & 64.8  & 91.4  & 96.3  & 79.7  & 96.4  & 98.9  & 527.5 \\
R &DSRN       & 64.5	&90.8	&95.8 &78.3	&95.7	&98.4 & 523.5 \\	
R &DIME(i2t)   & 63.0  & 90.5  & 96.2  & 77.9  & 95.9  & 98.3  & 521.8  \\
R &DIME(t2i)   & 62.3  & 90.2  & 95.8  & 77.2  & 95.5  & 98.5  & 519.5  \\
R &VSE+2AD  & \textbf{67.5}  & \textbf{93.6}  & \textbf{97.7}  & \textbf{81.3}  & \textbf{96.7}  & \textbf{99.2}  & \textbf{536.0}  \\ \cdashline{1-9}[1pt/1pt]
RG &VSE$\infty$  & 68.1  & 92.9  & 97.2  & 82.2  & 97.5  & \textcolor{blue}{\textbf{99.5}}  & 537.4 \\
RG &VSE+2AD       & \textcolor{blue}{\textbf{71.9}}  & \textcolor{blue}{\textbf{94.3}}  & \textcolor{blue}{\textbf{98.3}}  & \textcolor{blue}{\textbf{84.2}}  & \textcolor{blue}{\textbf{98.5}}  & 99.4  & \textcolor{blue}{\textbf{546.6}}  \\ \cdashline{1-9}[1pt/1pt]
RGP &VSE+2AD & \textcolor{red}{\textbf{72.5}}  & \textcolor{red}{\textbf{94.8}}  & \textcolor{red}{\textbf{98.7}}  & \textcolor{red}{\textbf{85.4}}  & \textcolor{red}{\textbf{98.9}}  & \textcolor{red}{\textbf{99.2}}  & \textcolor{red}{\textbf{549.5}}  \\
\bottomrule
\end{tabular}
\begin{tabular}{ccc:ccc:c}
\toprule
\multicolumn{6}{c}{Flickr30K 1K Test} &  \\ \cline{1-6} 
\multicolumn{3}{c}{Image Retrieval} & \multicolumn{3}{c}{Caption Retrieval} & \multicolumn{1}{l}{} \\
R@1  & R@5  & R@10  & R@1  & R@5  & R@10  & RSUM \\ \hline
\multicolumn{7}{l}{} \\ 
45.7  & 73.6  & 81.9 & 62.2  & 86.6  & 92.3  & 442.3\\
51.2       & 80.4  & 87.2  & 69.6  & 90.3  & 95.6  & 474.3  \\
-  & -  & -  & -  & -  & -  & - \\
54.7       & 82.2  & 88.6  & 70.5  & 88.0  & 92.7  & 476.7  \\ 
56.4       & 83.4  & 89.9  & 76.5  & 94.2  & 97.7  & 498.1  \\
43.9       & 74.2  & 82.8  & 67.9  & 89.0  & 94.4  & 452.2  \\
45.8       & 74.4  & 83.0  & 61.8  & 87.5  & 93.7  & 446.2  \\
52.8       & 79.0  & 87.9  & 70.1  & 91.6  & 97.2  & 478.6  \\
53.9       & 79.4  & 87.2  & 74.1  & 93.0  & 96.6  & 484.2  \\ 
\textbf{58.0}       & \textbf{85.0}  & \textbf{91.2}  & \textbf{76.9}  & \textbf{94.4}  & \textbf{98.2}  & \textbf{503.7}  \\ \cdashline{1-7}[1pt/1pt]
60.8  & 86.3  & 92.3  & 80.7  & 96.4  & 98.3  & 514.8  \\
\textcolor{blue}{\textbf{62.2}}  & \textcolor{blue}{\textbf{86.8}}  & \textcolor{blue}{\textbf{93.1}}  & \textcolor{blue}{\textbf{82.2}}  & \textcolor{blue}{\textbf{97.1}}  & \textcolor{blue}{\textbf{98.8}}  & \textcolor{blue}{\textbf{520.2}}  \\ \cdashline{1-7}[1pt/1pt]
\textcolor{red}{\textbf{63.5}} & \textcolor{red}{\textbf{87.6}}  & \textcolor{red}{\textbf{93.4}}  & \textcolor{red}{\textbf{83.1}} & \textcolor{red}{\textbf{97.7}}  & \textcolor{red}{\textbf{99.1}}  & \textcolor{red}{\textbf{524.4}}  \\ \hline \hline
\multicolumn{7}{l}{}              \\
45.6  & 76.4 & 84.4  & 63.4  & 87.2  & 92.7  & 449.7  \\
61.4  & 85.9  & 91.5  & 81.7  & 95.4  & 97.6  & 513.5 \\
59.2       & 86.0  & 91.9  & 77.8  & 95.1  & 97.6  & 507.6  \\
64.6       & 85.5  & 91.0  & 77.5  & 93.5  & 97.4  & 504.0  \\
60.1       & 85.5  & 91.8  & 77.4  & 95.0  & 97.4  & 507.2  \\
\textbf{59.1}       & \textbf{90.3}  & \textbf{93.5}  & \textbf{81.8}  & \textbf{96.1}  & \textbf{98.4}  & \textbf{524.7}  \\ \cdashline{1-7}[1pt/1pt]
66.7       & 89.9  & 94.0  & 85.3  & 97.2  & 98.9  & 532.0 \\
\textcolor{blue}{\textbf{69.2}}       & \textcolor{blue}{\textbf{91.3}}  & \textcolor{blue}{\textbf{95.6}}  & \textcolor{blue}{\textbf{87.1}}  & \textcolor{blue}{\textbf{97.9}}  & \textcolor{blue}{\textbf{99.3}}  & \textcolor{blue}{\textbf{540.4}}  \\ \cdashline{1-7}[1pt/1pt]
\textcolor{red}{\textbf{71.4}} & \textcolor{red}{\textbf{92.0}}  & \textcolor{red}{\textbf{95.8}}  & \textcolor{red}{\textbf{88.2}}  & \textcolor{red}{\textbf{98.4}}  & \textcolor{red}{\textbf{99.5}}  & \textcolor{red}{\textbf{545.3}}  \\
 \bottomrule
\end{tabular}}
\caption{\textbf{Cross-modal retrieval results} on MS-COCO and Flickr30K datasets. The top half of the table uses BiGRU as the text encoder; the bottom half is BERT. {VF} denotes vision feature, and R, G, and P mean {region}, {grid} and {patch} respectively. The best results are marked {bold} in \textbf{black}, \textcolor{blue}{\textbf{blue}} and \textcolor{red}{\textbf{red}} for region feature (R), ensemble of region$+$grid features (RG) and ensemble of region$+$grid$+$patch features (RGP) respectively. VSE+2AD is our proposed model, which enhances the VSE model by using \adcap for aggregation and \adcto for optimization. \label{tab:basic_res}}
\end{table*}

\subsection{Implementation Details}
\label{sec:details_extrs}
We implement our models using the PyTorch library. The dimension of the shared embedding space $d$ is 1024. We use the Adam optimizer with a mini-batch size of 128 and train our models with 25 epochs. The learning rate is set to 5e-4 with a decaying factor of 10\% for every 15 epochs.

\textbf{Visual Encoders}. We use Faster-RCNN \cite{ren2015faster} (ResNet-101 is pre-trained on ImageNet and Visual Genome) to extract \emph{region} feature directly \cite{anderson2018bottom}, and fine-tune it further with \emph{grid} feature (resolution $=$ 512$\times$512) \cite{jiang2020defense} before using it as a grid feature extractor. For the \emph{patch} feature, we fine-tune the pre-trained Vision Transformer \footnote{\href{https://huggingface.co/google/vit-base-patch16-224}{ vit-base-patch16-224}} \cite{dosovitskiy2020vit} using images with a resolution of 224$\times$224 to use it as a patch feature extractor.

\textbf{Text Encoders}. We experiment with BiGRU \cite{faghri2018vse++} and BERT-base\footnote{\href{https://huggingface.co/bert-base-uncased}{ bert-base-uncased}} \cite{devlin-etal-2019-bert}. Additional implementation and training details are given in the Appendix.





%

\subsection{Main Results}
Table~\ref{tab:basic_res} presents the full results over both image and caption retrieval and across two datasets (MS-COCO and Flickr30K). Results for MS-COCO is an average of over 5-folds of 1K test images. Our method is ``VSE+2AD'', which enhances the standard VSE model by introducing \adcap and \adcto. The top half of the table presents results where we use BiGRU as the text encoder, and the bottom half uses BERT.  
For models where we combine visual features from multiple visual encoders (e.g.\ ``RG'' which combines region and grid feature), we do so by simply taking the average similarity values to rank the candidates. 

Looking at the results, VSE+2AD (our model) outperforms almost all baselines/benchmark models consistently. Our model displays consistent improvement over the state-of-the-art method VSE$\infty$ \cite{chen2021learning} with the same visual (region feature by BUTD \cite{anderson2018bottom}) and text encoders (BiGRU). In particular, it obtains 1.4\% (520.5 $\rightarrow$ 527.8) and 1.0\% (498.1 $\rightarrow$ 503.1) relative gains on RSUM for MS-COCO and Flickr30K datasets. Such improvements are stable no matter using which combination of visual and text encoders.  We also see that combining visual encoders (``RG'' vs.\ ``R'') further boosts its performance (like RSUM from 527.8 to 534.4 for the MS-COCO dataset), and utilizing all types of visual features (``RGP'') produces the best performance (539.1 $\textgreater$ 534.4 $\textgreater$ 527.8).

\subsection{Comparison with pre-trained models}
\label{sec:compare-pretrained}
We next compare VSE+2AD (with BERT as the language encoder) to pre-trained vision language models: ViLBERT \cite{lu2019vilbert}, UNITER \cite{chen2019uniter}, OSCAR \cite{li2020oscar}, ALIGN \cite{Jia2021ScalingUV}, CLIP \cite{radford2021learning} and MVP \cite{li2022mvp} in Table~\ref{tab:comp_vls}.
These results are evaluated on the COCO 5K test images.

Without using any large-scale corpus for pre-training, our ensemble VSE+2AD$^{\text{RGP}}$ (that combines region/grid/patch features) outperforms two out of six pre-trained VL methods and is not much worse than OSCAR, even though it does not use any cross-modal interaction. Our model is substantially faster than these pre-trained models: our slowest ensemble model is still an order of magnitude faster. As for our method we can pre-compute and cache the visual and text features, so during retrieval the only operations needed are similarity calculation and ranking. Overall, these results demonstrate that our model strikes a good balance between performance and efficiency.

\begin{table}[t]
\centering
\small{
\renewcommand\arraystretch{1.1}
\setlength\tabcolsep{1pt}
\begin{tabular}{lcc:cc:cc}
\toprule
   & \multicolumn{4}{c}{COCO 5K Test} & &  \\ \cline{2-5}
 & \multicolumn{2}{c}{Image Retrieval} &  \multicolumn{2}{c}{Caption Retrieval} & & \\
Method  & R@1   & R@5   & R@1   & R@5   & RSUM  & \#OIPs  \\ \hline
ViLBERT & 38.6  & 68.2  & 53.5  & 79.7  & 406.9 & -      \\
UNITER  & 48.4  & 76.7  & 63.3  & 87.0  & 454.4 & 9.6$\times$   \\
OSCAR   & 54.0  & 80.8  & 70.0  & 91.1  & 479.9 & 4.3$\times$   \\
ALIGN   & 59.9  & 83.3  & 77.0  & 93.5  & 500.4 & 1.0$\times$     \\
CLIP   & 58.7  & 83.6  & 76.8  & 94.0  & 500.9 & -      \\
MVP    & 60.1  & 84.0  & 77.3  & 93.6  & 502.6 & - \\ \cdashline{1-7}[1pt/1pt]
VSE+2AD  & 52.5  & 80.2  & 69.5  & 91.2  & 475.9 & 45.3$\times$  \\
\bottomrule
\end{tabular}}
\caption{Comparison between VSE+2AD and pre-trained models. \#OIPs denotes operating items per second, and larger is better. The language encoder for VSE+2AD is BERT. And the shown number is the ensemble results considering \emph{region$+$grid$+$patch} visual features.
\label{tab:comp_vls}}
\end{table}

\begin{table}[t]
\centering
\small{
\renewcommand\arraystretch{1.1}
\setlength\tabcolsep{1pt}
\begin{tabular}{lccc:ccc:c}
\toprule
 & \multicolumn{7}{c}{COCO 5K Test}   \\ \cline{2-7}
 & \multicolumn{3}{c}{Image Retrieval} & \multicolumn{3}{c}{Caption Retrieval} & \\
Method    & R@1   & R@5   & R@10  & R@1   & R@5   & R@10  & RSUM  \\\hline
VSE+2AD    & 41.4  & 72.3  & 83.8  & 54.9  & 83.0  & 91.5  & 426.9 \\ \cdashline{1-8}[1pt/1pt]
$-$\emph{tok} \adcap  & 39.5  & 71.3  & 82.3  & 53.5  & 82.2  & 91.3  & 420.1 \\
$-$\emph{emb} \adcap  & 40.1  & 72.3  & 83.2  & 53.7  & 82.5  & 91.3  & 423.1 \\
$-$\adcap      & 38.5  & 72.4  & 81.8  & 53.3  & 82.0  & 91.1  & 419.1 \\ \cdashline{1-8}[1pt/1pt]
$-$\adcto     & 39.2  & 71.1  & 81.5  & 53.4  & 82.2  & 90.5  & 417.9 \\
\bottomrule
\end{tabular}}
\caption{Ablation results where we measure the \emph{token-level}, \emph{embedding-level}, overall \adcap pooling methods and \adcto optimization strategy. The \emph{region} feature with simple projection as visual encoder and BiGRU as the textual encoder. \label{tab:ablation}}
\end{table}

\subsection{Ablation Study}
\label{sec:ablation-study}
To understand the impact of \adcap (which improves feature aggregation in Section~\ref{sec:pooling}) and \adcto (which improves optimization in Section~\ref{sec:optimization}), we perform several ablation studies based on the COCO 5K test. For this experiment (in Table \ref{tab:ablation}), we use only the \emph{region} feature and BiGRU as the text encoder. Looking at the aggregate RSUM performance, we see both the token-level pooling (``$-$\emph{tok} \adcap''; Section \ref{sec:object-pooling}) and embedding-level pooling (``$-$\emph{emb} \adcap''; Section \ref{sec:embedding-pooling}) appear to be useful, although token-level pooling is arguably more important. In the case where we remove \adcap entirely (``$-$\adcap'') and use MeanPool as the aggregation method, the performance drops even more, suggesting complementarity. \adcto is the most impactful method, where taking it out produces the worst performance. To understand its impact qualitatively, we also look at the training curve of a VSE model trained with and without \adcto (Appendix \ref{verify_a2} Figure~\ref{fig:ab}). Here it is clear that \adcto is particularly useful in the early stages of training, where it helps the model to converge much faster. This earlier convergence ultimately impacts their final performance, where the VSE trained with \adcto reaches a plateau that is higher than the one without \adcto.

We next investigate the impact of \adcap further, by replacing it with other more advanced pooling strategies. As before, we use the region feature, BiGRU for text encoder, and COCO 5K test.
Table \ref{tab:appd_pool} presents the results.  Here we see that \adcap outperforms existing pooling strategies (best RSUM), even against more complex aggregators such as Seq2Seq, GCN and SelfAttn. It is also reasonably fast (competitive with other methods). These results once again highlight that our proposed pooling method has both performance and speed as we saw in Section \ref{sec:compare-pretrained}.

\begin{table}
\centering
\small{
\renewcommand\arraystretch{1.1}
\setlength\tabcolsep{0pt}
\begin{tabular}{lcc:cc:ccc}
\toprule
 & \multicolumn{4}{c}{COCO 5K Test} & & \\ \cline{2-5}
& \multicolumn{2}{c}{Image Retrieval}   & \multicolumn{2}{c}{Caption Retrieval}  & & \\
Aggregator        & R@1  & R@5 & R@1 & R@5 & RSUM &\#OIPs \\ \hline
LIP      & 38.4 & 69.3   & 52.2   & 81.1   & 414.8 &4.6$\times$\\
Seq2Seq  & 38.2 & 68.9   & 52.1   & 80.9   & 415.7 &1.8$\times$\\
GCN      & 40.7 & 71.3   & 53.5   & 81.5   & 419.2 &1.0$\times$\\
SelfAttn & 40.9 & 71.3   & 53.8   & 82.2   & 420.1 &1.6$\times$\\
AdaPool  & 40.2 & 71.8   & 52.8   & 81.4   & 418.3 &4.1$\times$\\
SoftPool & 39.2 & 69.8   & 52.8   & 81.4   & 419.8 &5.1$\times$\\
GPO      & 41.2 & 71.1   & \textbf{55.6}   & 82.0   & 422.9 &4.3$\times$\\
Manual       & 38.7 & 71.6   & 53.7   & 82.1   & 419.3 &\textbf{5.6}$\times$\\ \cdashline{1-8}[1pt/1pt]
\adcap       & \textbf{41.4} & \textbf{72.3}   & 54.9   & \textbf{83.0}   & \textbf{426.9} &4.9$\times$ \\ \bottomrule
\end{tabular}}
\caption{Performance of VSE+2AD using different aggregators. \#OIPs denotes operating items per second, and larger is better. Note that our proposed \adcap aggregation method is only slower than simple pooling, 5.6$\times \textgreater$ 4.9 $\times(\adcap) \textgreater$ others, but it no need for fussily manual tuning. \label{tab:appd_pool}}
\end{table}

\begin{figure*}[b]
\centering
\subfigure[Image Retrieval]{
    \label{fig:ir}
    \begin{minipage}[t]{0.52\linewidth}
		\includegraphics[width=1\linewidth]{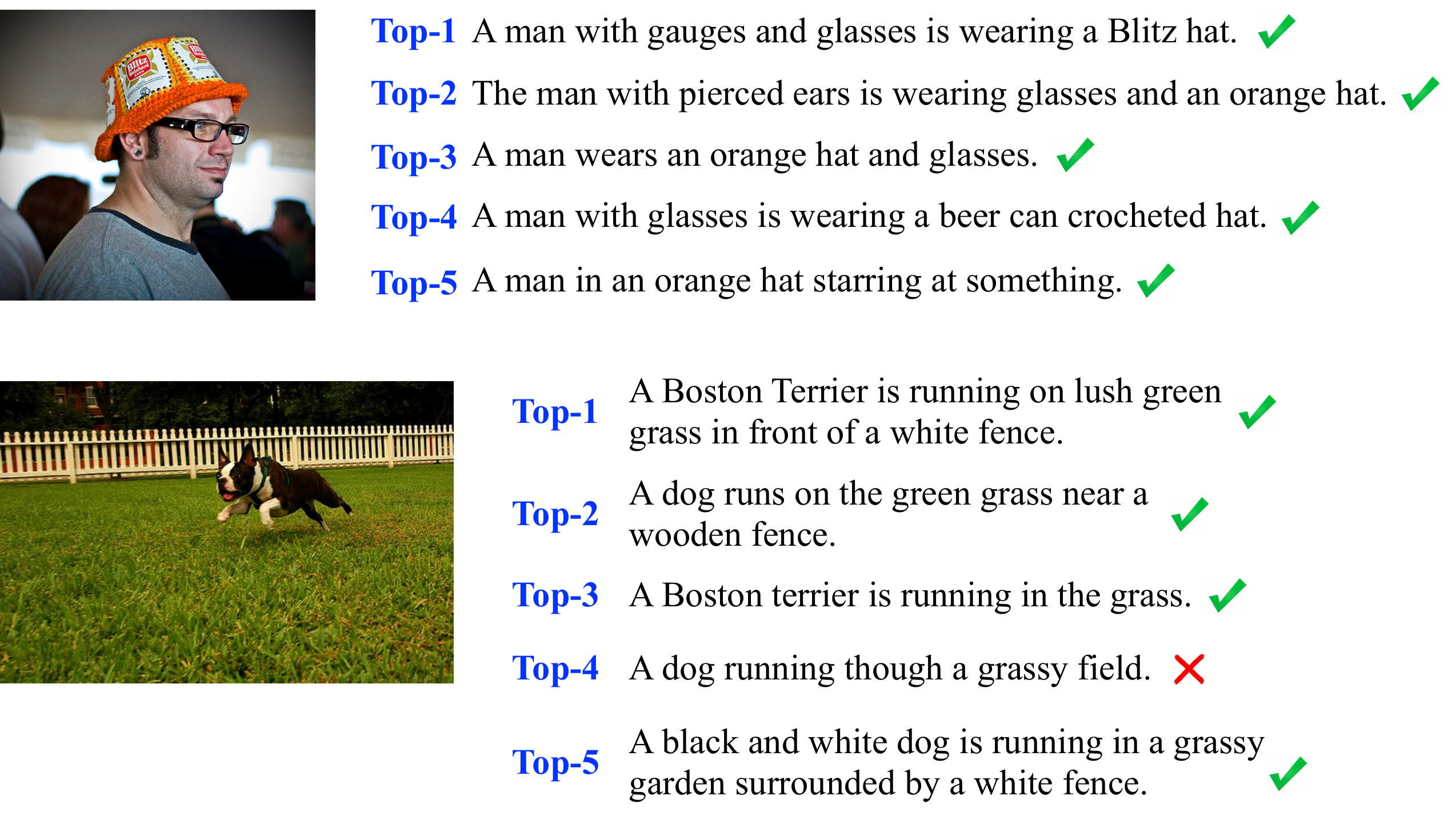}
	\end{minipage}}
\hspace{8mm}
\subfigure[Caption Retrieval]{
    \label{fig:cr}
    \begin{minipage}[t]{0.4\linewidth}
		\includegraphics[width=1\linewidth]{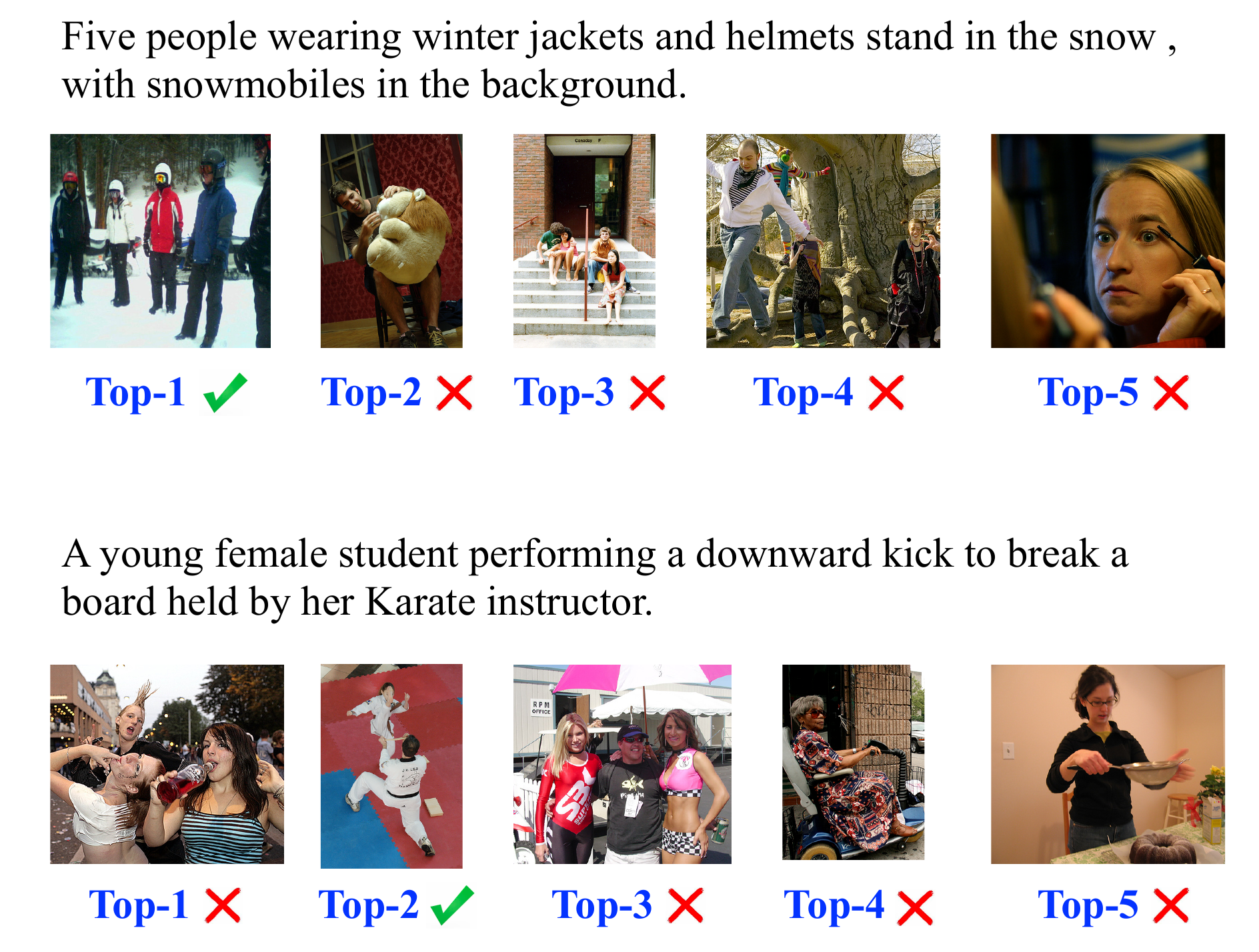}
	\end{minipage}}
\caption{Top-5 image retrieval and caption retrieval results on Flickr30K. The ground-truth results are marked with greed \textcolor{green}{$\checkmark$}, and the wrong results are indicated by red \textcolor{red}{$\usym{2718}$}.}
\end{figure*}

\subsection{Case Study}
\label{sec:case-study}
To validate the effectiveness of VSE+2AD, we present two examples for image retrieval and caption retrieval in Figure~\ref{fig:ir} and Figure~\ref{fig:cr}, respectively. As we can see in Figure~\ref{fig:ir}, the ``incorrect'' sentence retrieved by our model seems sensible, suggesting that this is likely noise in the data. Figure~\ref{fig:cr}, on the other hand, shows some genuine erroneous images retrieved by our model, and we suspect this is because it is a particularly difficult example where the caption is very descriptive and the details are difficult to be captured by VSE's bi-encoder approach.

\section{Conclusion}
\label{conclude}
In this paper, we propose methods to improve VSE's feature aggregation and optimization objective. For the former, we introduce a way to parameterize the aggregation function to allow the visual and text encoders to learn the best way to combine their features to produce a fixed-size embedding. For the latter, we propose a method that dynamically selects many negative samples that allows the VSE to converge faster with a better final performance. We compare our enhanced VSE model to several baselines and state-of-the-art models over two public datasets and demonstrate that it marries both performance (state-of-the-art retrieval results) and efficiency (orders of magnitude faster than pretrained models). As our proposed method is more suitable for practical application.


\clearpage
\bibliography{anthology,custom}
\bibliographystyle{acl_natbib}

\clearpage
\appendix
\setcounter{table}{0}
\setcounter{figure}{0}

\section*{Appendix}
\label{sec:appendix}


\section{Verification of Assumptions}
\subsection{Simple pooling strategy works best}
\label{verify_a1}
Although various complex methods (described in Section~\ref{related-work}) are explored for aggregation, we find that the simple pooling strategy works no worse than those complex methods by numerous experiments. It needs to be carefully manually tuned, like 5-MaxPool for visual feature and MeanPool for text feature. The results are shown in Table~\ref{verify_a1}-\ref{tab:appendix_pool}, where the similar conclusion is also verified in VSE$\infty$ \cite{chen2021learning}.

\begin{table}[ht]
\centering
\small{
\renewcommand\arraystretch{1.1}
\setlength\tabcolsep{0pt}
\begin{tabular}{lcc:cc:ccc}
\toprule
 & \multicolumn{4}{c}{COCO 5K Test} & & \\ \cline{2-5}
& \multicolumn{2}{c}{Image Retrieval}   & \multicolumn{2}{c}{Caption Retrieval}  & & \\
Aggregator        & R@1  & R@5 & R@1 & R@5 & RSUM \\ \hline
LIP      & 38.4 & 69.3   & 52.2   & 81.1   & 414.8 \\
Seq2Seq  & 38.2 & 68.9   & 52.1   & 80.9   & 415.7 \\
GCN      & 40.7 & 71.3   & 53.5   & 81.5   & 419.2 \\
SelfAttn & 40.9 & 71.3   & 53.8   & 82.2   & 420.1 \\
AdaPool  & 40.2 & 71.8   & 52.8   & 81.4   & 418.3 \\
SoftPool & 39.2 & 69.8   & 52.8   & 81.4   & 419.8 \\
GPO      & 41.2 & 71.1   & \textbf{55.6}   & 82.0   & \textbf{422.9} \\ \hline
Manual   & 38.7 & \textbf{71.6}   & 53.7   & \textbf{82.1}   & 419.3 \\ \bottomrule
\end{tabular}}
\caption{Comparison of complex aggregators and manually chosen pooling method evaluated with MS-COCO 5K. \label{tab:appendix_pool}}
\end{table}

\subsection{Hardest triplet loss slows the convergence}
\label{verify_a2}
\begin{figure}[ht]
\centering
\includegraphics[width=\linewidth]{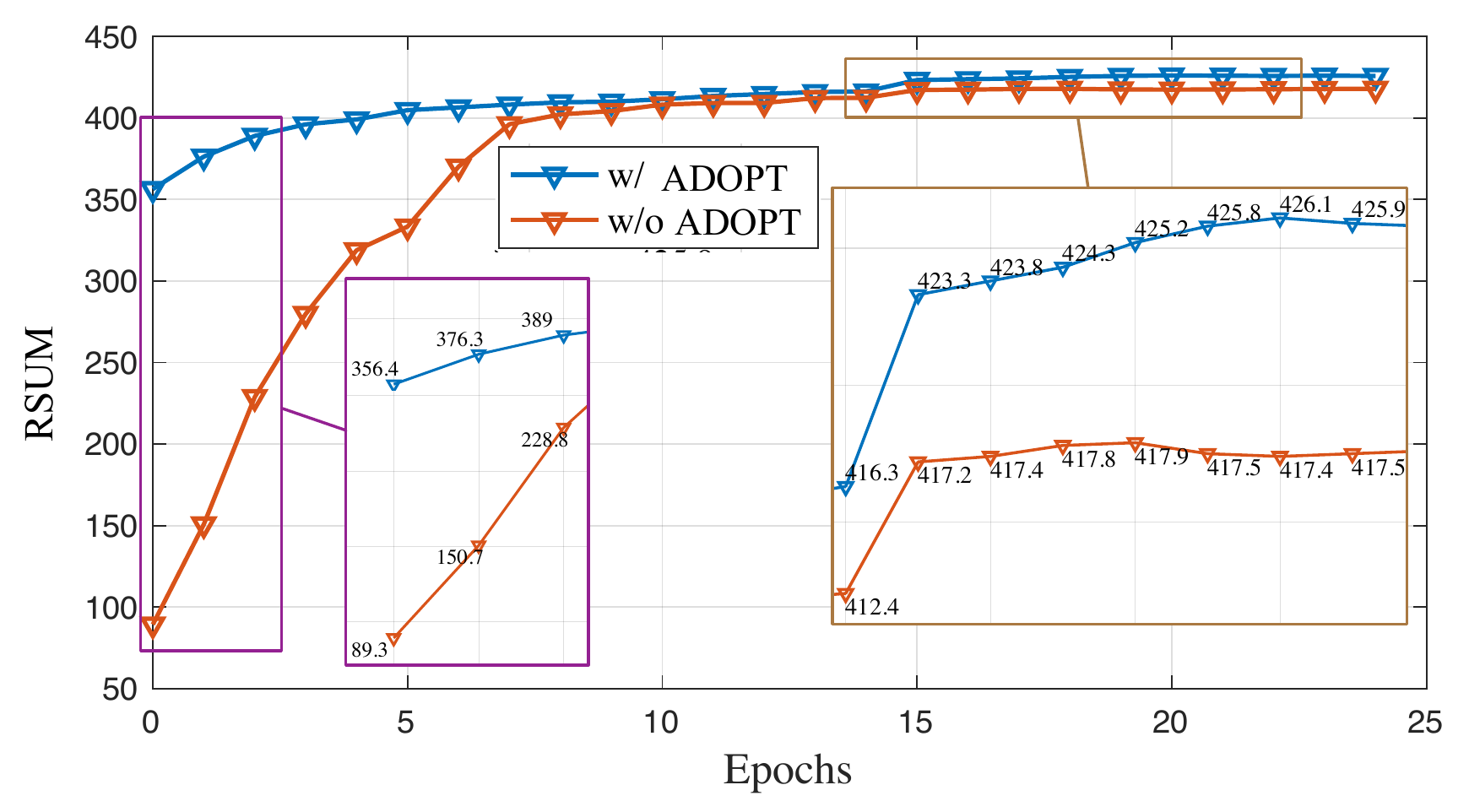}
\caption{Performance of VSE trained with and without \adcto. \label{fig:ab}}
\end{figure}

Figure~\ref{verify_a2}-\ref{fig:ab} shows the comparison of VSE model with and without our proposed optimization objective, \adcto. Note that hard triplet loss \cite{faghri2018vse++} is used for optimization when \adcto is not used. The optimization target \adcto can fit the model more quickly, thus further improving the potential of the model, that is performance is better in the latter stage.

\section{Additional Implementation Details}
\label{sec:add_implements}

\textbf{Basic settings}. The margin $\alpha$ for hard triplet loss is set to 0.2 and the $\tau$ used in InfoNCE is 0.05. The image and text features extracted from the encoders use L2 normalization. The common learning rate is set to 5e-4, while the learning rate of the pre-trained modules (like BERT, ResNet and ViT) is 10\% of its.

\textbf{Vision Encoders}. When using \emph{region} feature that directly extracted from BUTD \cite{anderson2018bottom}, the multilayer perceptron is used to map the visual feature dimension into 1024 as the same as text feature dimension. For \emph{grid} feature, the warm-up strategy is used for the first epoch. Then, all parameters are optimized in the rest of 24 epochs. For \emph{patch} feature, the original image is changed to 224$\times$224 resolution with the size of a patch as 16.

\textbf{Language Encoders}. When using BiGRU as the backbone, the token dimension is 300 and the hidden dimension is 1024. Only one layer is considered and the bidirectional features are averaged as the output feature. For BERT, the hidden dimension is 768 and the multilayer perceptron is also used to map the text feature dimension into 1024 as the same as visual feature dimension.

\section{Additional Experiments and Results}
\label{sec:add_exp_res}

\begin{figure}[]
    \centering
    \includegraphics[width=\linewidth]{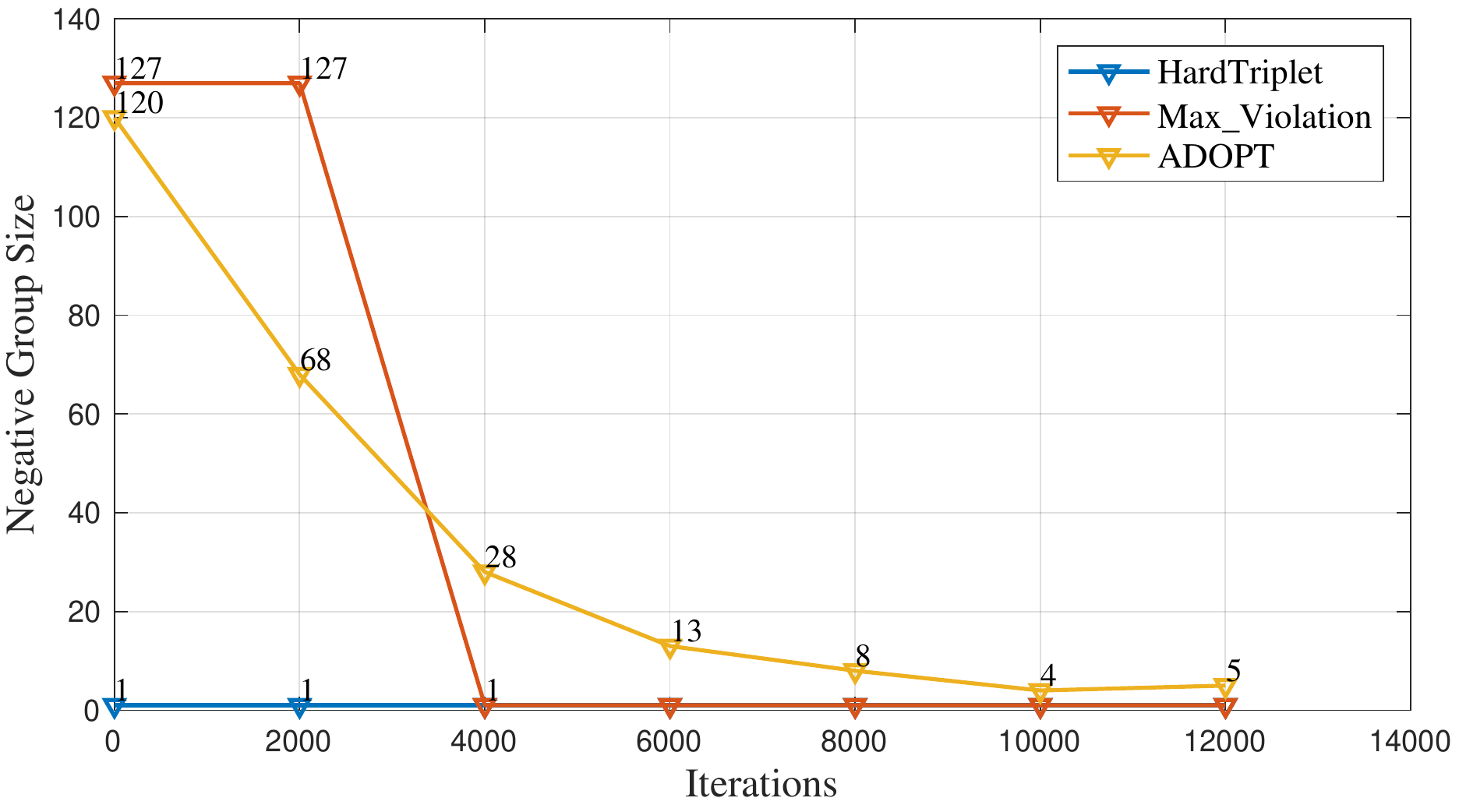}
    \caption{Visualization of negative group size learned by \adcto on the first 5 epochs (about 12000 iterations). \label{fig:appd_negsize}}
\end{figure}

We present Figure~\ref{sec:add_exp_res}-\ref{fig:appd_negsize} to show how the size of the negative samples changes during training with \adcto. We can see that \adcto starts with a large number of negative samples, but that decreases over time to only 4--5 samples at the end of the 5$^{th}$ epoch (yellow line). Different from the common hard triplet optimization that only considers one hardest negative sample, the max\_violation strategy \cite{faghri2018vse++} considers the rest samples within the same mini-batch as negative samples only on the first epoch and the rest epochs are the same as the hard triplet loss.

We last investigate the impact of the balance module in \adcap (Equation \ref{eqn:balance}). Table~\ref{sec:add_exp_res}-\ref{tab:appd_balance} shows that manually tuned weights underperform substantially compared to automatically learned weights\footnote{For ``random'', we select the weights randomly and run them 5 times and compute the average to reduce variance.}. 

\begin{table}
\centering
\small{
\renewcommand\arraystretch{1.1}
\setlength\tabcolsep{0pt}
\begin{tabular}{cccc:cc:c}
\toprule
 & & \multicolumn{4}{c}{COCO 5K Test} &  \\ \cline{3-6}
\multicolumn{2}{c}{Weights} & \multicolumn{2}{c}{Image Retrieval} & \multicolumn{2}{c}{Caption Retrieval} &  \\ 
\emph{obj}  & \emph{emb}  & R@1  & R@5  & R@1  & R@5  & RSUM  \\ \hline
0.25        & 0.75        & 39.8 & 70.2 & 51.9 & 80.4 & 416.6 \\
0.5         & 0.5         & 40.1 & 70.6 & 52.8 & 81.1 & 418.2 \\
0.75        & 0.25        & 40.6 & 71.4 & 54.1 & 82.2 & 421.3 \\
random      & random      & 39.3 & 69.4 & 51.2 & 80.1 & 413.9 \\ \cdashline{1-7}[1pt/1pt]
\adcap      & \adcap      & 41.4 & 72.3 & 54.9 & 83.0 & 426.9 \\ \bottomrule
\end{tabular}}
\caption{Different choices of parameters for fusing token-level and embedding-pooling. \label{tab:appd_balance}}
\end{table}

\end{document}